\documentclass[journal,10pt,twocolumn]{IEEEtran}

\usepackage{amssymb, amsmath, amsthm}
\usepackage{amsfonts, bbm}
\usepackage{graphicx}
\usepackage{epstopdf}
\usepackage{times}
\usepackage{float} 
\usepackage{array}
\usepackage{arydshln}
\usepackage{bibentry}
\usepackage{booktabs}

\usepackage{cite}
\usepackage{subfigure}
\usepackage{enumitem}
\usepackage{color}
\usepackage{setspace}
\usepackage{bigstrut}
\usepackage{multirow}
\usepackage{algorithm}
\usepackage{algorithmic}
\usepackage{subfig}
\usepackage{mathtools}

\newcommand{\figref}[1]{Fig.~\ref{#1}}
\begin{document}

\title{Multi-tier Drone Architecture for 5G/B5G Cellular Networks:  Challenges, Trends, and Prospects}

\author{Silvia Sekander, Hina Tabassum,  and Ekram Hossain\thanks{The authors are with the Department of Electrical and Computer Engineering, University of Manitoba, Canada (Emails: sekandes@myumanitoba.ca, \{Hina.Tabassum, Ekram.Hossain\}@umanitoba.ca). This work was supported by the Natural Sciences and Engineering Research Council of Canada (NSERC).}
}

\maketitle

\begin{abstract}

Drones (or unmanned aerial vehicles [UAVs]) are expected to be an important component of fifth generation (5G)/beyond 5G (B5G) cellular architectures that can potentially facilitate wireless broadcast or point-to-multipoint transmissions.
The distinct features of various drones such as the maximum operational altitude, communication, coverage, computation, and endurance impel the use of a multi-tier architecture for future drone-cell networks. In this context, this article focuses on investigating the feasibility of multi-tier drone network architecture over traditional single-tier drone networks and identifying the scenarios in which drone networks can potentially complement the traditional RF-based terrestrial networks. We first identify the challenges associated with multi-tier drone networks as well as drone-assisted cellular networks. We then review the existing state-of-the-art innovations in drone networks and drone-assisted cellular networks.  We then investigate the performance of a  multi-tier drone  network in terms of spectral efficiency of downlink transmission while illustrating the optimal intensity and altitude of drones in different tiers numerically.  Our results demonstrate the specific  network load conditions (i.e., ratio of user intensity and base station intensity) where deployment of drones can be beneficial (in terms of spectral efficiency of downlink transmission) for conventional terrestrial cellular networks.
\end{abstract}


\section*{Introduction}

The role of unmanned aerial vehicles (UAVs) with wireless communications functionalities (also termed as drones or aerial base stations [BSs])  is authoritative in fulfilling the broadcasting/point-to-point/point-to-multipoint communication requirements of 5G/B5G cellular networks~\cite{DSC}. 
For instance, typical wireless broadcasting applications include  satellite news-gathering, live sports coverage, portable field monitoring and video streaming.
Several projects  by the industry have already been initiated, e.g., Project Aquila by Facebook, cell-on-wheels and wings (COW-W), Google projects such as SKYBENDER  that are designed for drone-based Internet services~\cite{fb}.
%Historically, drones have been used for a wide variety of  applications such as  military, weather monitoring, forest fire detection, traffic control, cargo transport, emergency search and rescue work.  
%Drones have also  been considered as a  key substitute for terrestrial cellular networks in natural disaster scenarios. 
The primary benefits of drone technology are:~(i)~drones can operate in dangerous/disastrous environments,~(ii)~drones can be relocated easily and rapidly based on demand,~(iii)~drones can improve coverage  due to improved higher line-of-sight (LoS) connections with the ground users,~(iv) drones have adjustable height to meet Quality-of-Service (QoS) requirements based on user intensities, desired data rate, interference/blockage effects, etc.

Drones have the potential to substitute as well as complement the terrestrial cellular networks by serving users in severe shadowing or interference conditions, serving the overloaded or damaged terrestrial BSs,  serving users around idle BSs in ultra-dense networks, and users in rural areas.  In scenarios where it is not economically feasible or practically possible to install a new infrastructure immediately, drones can assist terrestrial networks by offloading the users either fully (in case of disasters) or partially (in case of overloaded/damaged BSs). In some cases where active users are very low compared to active BSs, deploying few drones  may allow a significant portion of BSs to be inactive; thus reducing  power consumption.

\begin{figure*}
\begin{center}
\includegraphics[scale=.5]{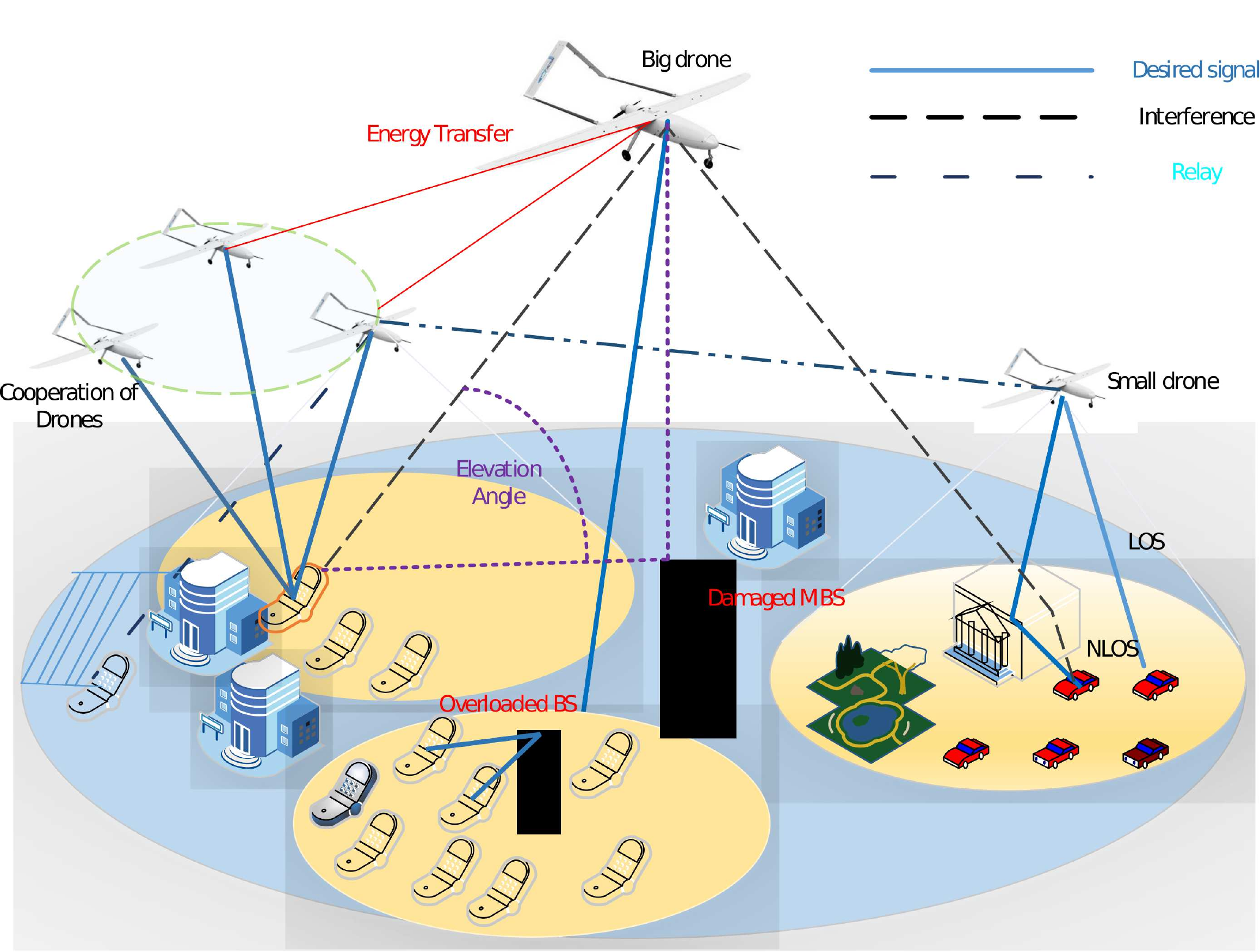}
\end{center}
\caption{A drone-assisted cellular network  where multiple tiers of drones exist and users are distributed homogeneously. The drones are connected
to terrestrial networks via a satellite link or an air-to-ground (AtG) wireless link. }
\label{sys}
\end{figure*}

Drones can be distinguished based on their size, weight, and power (SWAP) constraints.
The SWAP constraints directly impact the maximum operational altitude, communication, coverage, computation, and endurance capabilities of a drone~\cite{MAG}. 
For instance, low altitude platforms (LAPs)  have low power and low capacity both in terms of payload (ranging
from a few dozen grams 
to 5-7 kilograms) and autonomy (10 to 40 minutes depending on the battery capacity, mobility pattern, and payload weight). Due to their small form factor, LAPs can lift a
very limited weight and thus may be unable to carry LTE equipments (typical weight is 10~kg). 
%Subsequently, LAPs are typically deployed up to an altitude of  10~km~\cite{MAG}. 
Despite the aforementioned drawbacks, LAPs are more
cost-effective and can be swiftly
deployed. Besides, short-range LoS communication
links can be established efficiently to enhance coverage.  
%Moreover, LAPs are typically made for a single purpose and thus needs fewer number of chips leading to reduced maintenance cost.
Compared to LAPs, higher altitude platforms (HAPs) provide wider coverage and longer endurance. As such, HAPs are preferred for reliable wireless coverage in very large geographic areas~\cite{WCU}.
%HAPs such as  X-45, predator B, Darkstar, and Global Hawk can operate in the stratosphere, i.e., over 10~km~\cite{MAG}. 

The distinct features and capabilities of different drones  give rise the following questions: 
\begin{itemize}
\item whether it is beneficial to form a communication network with different types of drones in multiple tiers (i.e., drones at different altitudes) instead of a single tier?
\item if yes, then what is the impact of chosen intensities and altitudes  of drones in different tiers? What should be the  optimal altitudes/intensities for drones at different tiers?
\item whether and in which scenarios, drones can add to the performance of existing terrestrial cellular networks? 
\end{itemize}
Motivated by the aforementioned queries, in this article, we first review existing achievements and innovations in drone networks and drone-assisted cellular networks. We then identify challenges associated with multi-tier drone networks as well as drone-assisted cellular networks. These include optimizing energy consumption of drones, interference management, energy-aware and interference-aware 
deployment of multi-tier drones (i.e., optimizing the number and placement of drones in different tiers, altitude of various tiers).

A typical drone-assisted cellular network  is shown in
Fig.~1.  We will investigate the feasibility  and significance of a  multi-tier drone  network over single-tier drones in terms of spectral efficiency of downlink transmission. Our results show that the optimal intensity of drones in different tiers vary in different environments such as high rise urban, sub-urban, and dense urban.  In addition, our results demonstrate the specific  network load condition where the deployment of drones can be beneficial (in terms of network throughput) for terrestrial cellular networks.

%The rest of the article is organized as follows. Section~II details the challenges associated with drone-assisted cellular networks whereas Section~III provides an overview of the existing  state-of-the-art to highlight the potential research gaps. Section~IV analyzes the feasibility of multi-tier drone network over single-tier drone network and conventional terrestrial networks. Finally, Section~V concludes the paper.

\section*{Potential Challenges Associated with Multi-Tier Drones}
The deployment of drones or multi-tier drones can potentially take over the cellular transmissions in disastrous situations. Moreover, integrating drones to overloaded terrestrial networks may also offer benefits by offloading traffic, reducing handovers for highly mobile users, etc. Nonetheless, several challenges that need to be tackled  prior to harnessing the benefits of drone networks are discussed in the following. 

\subsection*{Deployment of Drone-Aided Cellular Network  and Air Traffic Control Systems for Drones}
The  user traffic is expected to grow significantly in future cellular networks thus leading to overloaded situations. At the same time, the terrestrial BSs are also expected to grow massively thus remedying the overloaded situations. 
It is therefore crucial to deploy drones opportunistically keeping in view the 
user load as well as the active BSs. Note that turning off terrestrial BSs and offloading users to drones may not always be useful as the communication distance from users to drones is typically large. Therefore, determining the correct load situation (user to drone ratio) at which drones may positively assist cellular networks is crucial. In this regard, offloading mechanisms leading to spectral or energy efficiency enhancement can be designed for users  depending on their interferences/desired QoS and/or the traffic load of terrestrial BSs. Mobility of the drone-cells (in both the vertical and horizontal dimensions)
as well as the characteristics of the air-to-ground (AtG) channels need to be considered for optimal deployment of drones in the designated drone corridors or `air highways' for drones. A reliable and efficient air traffic control system will be also essential for deployment of drones.

\subsection*{Trajectory Planning and Mobility Control for Drones} 

Finding an optimal flying trajectory for a drone is a challenging task due to 
practical constraints such as availability of the air corridors, connectivity, fuel limitation, collision, and terrain avoidance. Also, altitude and coverage of drones is  limited by the regulations in regional borders, military and civil aviation.  One useful method is to model the dynamics of drones taking into account the position and velocity. 
For instance, adaptive communications can be jointly designed
with mobility control to improve the drone communications. For example, in favorable AtG channel conditions, drones may lower their speed to sustain improved wireless connectivity or drones may increase their speed in restricted environments. Drones can also cooperate to adapt to the mobility of the users to decrease handover, optimize power and resource allocations, and avoid collisions. For the typical cellular applications,
rotary-wing drones are deemed as suitable that hover above the coverage
area and in turn serve as static aerial BSs. In this case, no dedicated trajectory planning is needed.

\subsection*{Operational Altitude of Drones}
Due to SWAP constraints, different kind of drones may be restricted to different operational altitudes which may or may not be favorable for a given urban environment. For instance, users in high-rise urban scenarios may require higher LoS connectivity whereas users in sub-urban scenarios may need higher degree of path-loss reduction. Note that the higher altitude of drones promotes higher LoS connectivity since reflection and shadowing get diminished, whereas, lower altitude ensures reduction in  path-loss. Thus a  trade-off between LoS connectivity and path-loss exist that can be potentially handled by the deployment of multi-tier drones due to the flexibility of selecting different altitudes for drones.   

\subsection*{Terrestrial-Drone Interference}
The presence of various drones in a single tier or multiple tiers  significantly  impacts the desired communication between a given drone and its associated  user. Thus interference management, taking into account the AtG channel features and mobility, is crucial.  Interference can be mitigated by utilizing different frequency spectrum at different tiers, e.g., mm-wave for LAPs and RF for HAPs. It should be also noted that, mobility of drones creates Doppler shift, which causes severe inter-carrier interference  at higher transmission frequencies. 
For spectrum sharing scenarios, heights and intensities of various tiers of drones can be optimized to mitigate interference and maximize network throughput.
Coordination among different drones may also be established to reduce interference and enhance the capacity of a drone user. Nonetheless, the absence of fixed backhaul connection for drones makes the coordination among drones more challenging. Therefore, to enable efficient coordination among drones, high capacity backhaul links would be required.

\subsection*{Energy Consumption of Drones}

Integration of drones with the terrestrial networks may reduce unnecessary energy consumption of the terrestrial BSs. For example, if there are several  active BSs serving very few users, a single drone may take over and serve the active users while allowing a certain fraction of BSs to become idle; thus saving energy
and minimizing the interference in the terrestrial network. Nonetheless,   the energy consumption of drones may also be high depending on their mobility, transmission power, and  circuit power consumption. As such, the energy efficiency of the drone-assisted cellular network needs to be carefully optimized in order to select the appropriate number of active BSs and drones while considering the degree of  reduction in terrestrial interference and power consumption, the weather conditions, size of the area to be served, network load, or network utility. Also, the movement of the drones should be carefully controlled by taking into account the energy consumption associated with every maneuver (especially in ascending direction). In this context, energy-efficient mobility patterns can be designed considering appropriate energy consumption models for both the cases: when the drones "hover" only and when the drones move continuously.

\subsection*{Limited Endurance}
Endurance in drones is  problematic, especially in LAPs because of the 
SWAP limitations.    Since drones mainly rely on  rechargeable  battery  power  source,    energy  harvesting  at drones is crucial to
increase the  endurance time  without  
adding significant mass or  size 
of   the   fuel   system. 
Note that HAPs may be able to harvest solar  energy due to the
relatively large dimensions of the platform; however, this may not be possible at LAPs. Nonetheless, ambient RF energy harvesting can be utilized at LAPs to harvest energy from the transmission signals of neighboring LAPs or HAPs. Another alternative is dedicated wireless power transfer from HAPs to ensure a better quality of charging. However, this may increase  the resource consumption at HAPs significantly depending on the distance, environment, and required power of the LAP. As such, schemes for maximization of endurance time  should be designed keeping in view the aforementioned trade-offs.

\subsection*{Cost, Security and Lack of Regulations}
The cost of
deploying a drone network for wireless broadcasting/multi-point transmissions is a fundamental
challenge mainly due to the purchasing cost of drones, cost of  accessories of drones, and their
maintenance cost. Depending on the size and operation of the drones, the battery requirements
will also vary and thus the operating cost of a sophisticated drone can be much higher.
Also, in case of aerial collisions, the maintenance cost of drones may significantly increase.
%We also agree that the security can be a major concern. 
%For instance, drones might be used to perform attacks on the civilians. Unethical operators with an intention to
%commit genocide can use the drones to destroy society. 
Proper authentication, regulation, and
operational techniques are needed to reduce the possible disastrous and dangerous
situations. The Civil Aviation Authority (CAA) in the United
Kingdom (UK) gives legal guidance on the usage of drones in UK Airspace depending on
the weight (less than 20 kg). For small-sized UAVs, visual-LoS (VLoS) is acceptable without
permission, i.e., a maximum distance of 0.5 km horizontally and 100 m in altitude from the
remote pilot. For extended- and beyond-VLoS, the drones must be equipped with means to
avoid collisions, be visible to other airspace users, and resilience to meteorological conditions.
For larger drones, collision avoidance and resiliency to the effects of wind are mandatory.

\subsection*{Backhauling Cellular Communication}
Providing backhaul connections to  all BSs in  future cellular networks may be practically infeasible due  to  their  possible outdoor/remote/hard-to-reach locations. Existing backhaul solutions  rely  on either RF spectrum or  microwave  frequencies or wired alternatives  at  the  expense  of
severe interference, congestion, and higher cost. By exploiting the  LoS connectivity of drones (due to their adjustable heights) in backhauls, the high capacity backhaul links for terrestrial BSs can be established.  
However, to provide reliable backhaul connections, drones also need high capacity communication link  with terrestrial wireless backhaul hub. 
Thus
the operating cost of drones may be high. In this regard,  options such as unmanned
balloons (typically solar-powered) may  be investigated. In order to be cost-effective, drones can also opportunistically utilize technologies in the unlicensed radio spectrum such as free space optics (FSO) and millimeter wave (mmWave) in conjunction with those using the traditional RF spectrum.  

\begin{table*}[!htbp]
\caption{Summary of state-of-the-art of drone-assisted wireless communications} % title of Table
\footnotesize
\centering % used for centering table
%\scalebox{0.5}
%\resizebox{\columnwidth}{!}
{

\begin{tabular}{|p{0.6cm}| p{2.0cm}| p{1.4cm}| p{2cm}|p{3.5cm}| p{2cm}| c|} % centered columns (4 columns)

\hline\hline %inserts double horizontal lines

Ref & Objective & Mobility & Type of BSs & Solution approach & Number of drones \\ [0.5ex] % inserts table

%heading

\hline % inserts single horizontal line

\cite{OLA} & Altitude Optimization & Static & Drone-only & Optimization & Single drone\\ % inserting body of the table

\hline 

\cite{DSC} & Altitude optimization & Static & Drone-only & Optimization & Two drones\\

\hline 

\cite{OTT} & Power optimization, traffic offloading & Static/mobile & Drone-only & Facility location framework, optimal transport theory & Multiple drones\\

\hline 

\cite{UAH} & Traffic offloading & Mobile & Hybrid (Drone \& terrestrial) & Reverse neural model & Multiple drones\\

\hline 

\cite{UAVU} & Optimal placement & Static/mobile & Drone-only  & Stochastic geometry/disk covering problem  & Single drone\\

\hline 

\cite{SSDN} & Intensity optimization & Static  & Hybrid (drone \& terrestrial) & Optimization  & Multiple drones\\

\hline 

\cite{E3D} & Optimal placement & Static & Hybrid (drone \& terrestrial)  & Optimization  & Multiple drones\\

\hline 

\cite{ON3} & Intensity optimization & Static & Drone-only  & Particle swarm optimization   & Multiple drones\\

\hline 

\cite{TNF} & Optimal placement  & Static & Hybrid (Drone \& terrestrial)  & 3D placement algorithm   & Multi-tier drones\\
\hline

\cite{H} & Optimal placement  & Static & Hybrid (drone \& terrestrial) & Minimax facility problem   & Multiple drones\\
\hline

\cite{MIT} & Power optimization  & Static & Drone-only  & $K$-means clustering   & Multiple drones\\
\hline

\cite{ED} & Intensity optimization  & Static & Drone-only  & Circle packing theory   & Multiple drones\\
[1ex] % [1ex] adds vertical space

\hline %inserts single line

\end{tabular}

}

\label{table:nonlin} % is used to refer this table in the text

\end{table*}

%\subsection{Other Issues}
%
%
%
%\subsubsection{MIMO at Drones} Harnessing the spatial multiplexing gains of MIMO  in drones is generally not feasible due to the absence of rich scattering in AtG channel and complex hardware, signal processing, and power consumption requirements.
%
%
%\subsubsection{Mobility Control}
%
%Adaptive communications can be jointly designed
%with mobility control to improve
%the drone communications. For example, in good AtG channel conditions, drones may lower their speed to sustain improved
%wireless connectivity. Drones can also cooperate to adapt to the mobility of the users to decrease handover, optimize power and resource allocations, and avoid collisions.

%\subsubsection{Inter-Carrier Interference} Mobility of drones creates Doppler shift, which causes severe inter-carrier interference for RATs at higher transmission frequencies. 

\section*{Existing State-of-the-Art Techniques}
In   this   section,   we   provide   an   overview   of   the   recent  research  studies  that  have dealt  with  the performance optimization of drone-assisted cellular networks. We review  these studies
mainly based on their objectives such as altitude/location optimization, energy-efficiency optimization,  intensity optimization, etc.  A summary of the review is provided in Table~I where we compare various research studies in terms of their objectives/applications, their solution approach, and considered type and number of drones.

\subsection*{Altitude Optimization for Drones}
In \cite{OLA}, the authors  derive the optimal altitude of a UAV in order to maximize the coverage for users. It is shown that the optimal altitude of the UAV strongly depends on the statistical parameters of the underlying environment and  path-loss.  Shadowing and scattering caused
by the man-made structures are considered in addition to  free space path-loss (FSPL). The additional path-loss  has a
Gaussian distribution which is approximated by the mean value of the distribution~\cite{OLA}.
A closed-form expression is derived to estimate the LoS probability between the UAV and  the ground receiver which depends on the elevation angle and environment variables. 
\cite{DSC} investigates the  impact of altitude on the downlink coverage of a static UAV. The optimal UAV altitude is determined such that the power of the UAV can be minimized. Further, the optimal deployment (altitude and distance between UAVs) of two UAVs is determined considering both interference and interference-free situations  to maximize the coverage.

\subsection*{Optimization of 3-D Placement of Drones}
The authors in \cite{E3D} formulate a 3-D placement problem for the  UAVs to maximize the revenue of the network where revenue is proportional to the number of drone cell users. They  propose a bisection search algorithm which jointly determines the drone altitude along with the coverage area. 
Optimal placement of drones is also studied in \cite{TNF} in a  network where drones assist the macro cells by serving users in the time of congestion. A 3D-placement algorithm is used to effectively serve the cellular users. 
In \cite{H}, the authors study the optimal placement of cooperative UAVs 
 with a target of minimizing network delays. 
They formulate the problem as a min-max facility problem and 
assign the UAVs to specific demand areas. \cite{UAVU}  analyzes the performance for both static and dynamic UAVs. The optimal altitude of a static UAV is derived  to maximize the downlink sum-rate  considering a network where both D2D transmitters and drone users are distributed as Poisson Point Process (PPP).  Expressions for coverage probability and achievable rates are provided. Moreover, the number of stop points for mobile UAV is optimized to minimize the transmit power of the UAVs.

\subsection*{Optimization of  Drone Intensity}
The authors in \cite{SSDN} investigate the impact of spectrum sharing in a drone small cell (DSC) network coexisting with the traditional cellular BSs where the DSCs are placed at a limited height. 
The authors calculate the optimal DSC density to achieve maximum throughput of the DSCs. Another interesting study is \cite{ED} where the authors consider a network of multiple UAVs with directional antennas. For a given target area, they maximize the coverage performance by ensuring minimum required transmit power of the UAVs.  They investigate the density of the UAVs required for a specific altitude of the UAVs. With the increase in the UAV height, the number of drones to cover a certain target area decreases.
Another interesting study is \cite{ON3} where a 3-D placement problem for drones is formulated to serve a set of users in the downlink not covered by macrocells. Due to the complexity of the optimal placement policy,  particle swarm optimization (PSO) is used. 

\subsection*{Traffic Offloading/User Association }
The authors in \cite{UAH} study the UAV deployment problem focusing on traffic offloading in a heterogeneous network composed of macro and small cells. By developing a neural cost model, they determine the locations where UAVs need to be placed depending on user demand. They model the  cost as a function of coverage, capacity, delay and achievable LOS. The authors in \cite{OTT} also investigate the traffic offloading problem in a network where the locations of the UAVs are pre-determined. Using optimal transport theory (OTT), user associations are determined in order to minimize the total transmit power of the UAVs.  It is shown that an appropriate user association method can lead to a power-efficient UAV network. 

\subsection*{Power Minimization of  Drones}
The authors in \cite{OTT} investigate the power minimization problem in a UAV network.  Given a certain cell boundary, the locations of the UAVs are derived using a facility location framework in order to ensure minimum transmit power. In this facility location framework, given a set of clients and facilities, the optimal facility locations are determined to achieve minimum transportation cost. The transmit powers of the UAVs are considered as cost. It is shown that by adjusting the altitudes and locations of the UAVs,  minimum power consumption can be achieved.  
Using tools from stochastic geometry, 
in \cite{MIT}, the authors discuss an energy-efficient UAV deployment method to collect information from IoT devices in uplink. Using a $K$-means clustering approach, the ground devices are first clustered and one UAV is assigned to serve each cluster.  It is shown that the proposed deployment strategy reduces  power of the IoT devices by 56\% when compared to the classical stochastic geometry-based Voronoi deployment approach.

\subsection*{Summary}
Most of the existing research focus on investigating the deployment issues such as optimizing the intensity/altitudes  of drones in a single-tier drone network. None of the studies analyzes the feasibility of a multi-tier drone network  where different drone tiers may assist users by providing a trade-off between LoS connectivity and path-loss. Note that the LoS connectivity increases with increasing altitude and/or density of drones since the path-loss components due to reflection and shadowing get diminished. On the other hand,  increasing the altitude results in an increased distance from the user and hence decreases the received signal strength. As such, the multi-tier drones may provide plausible gains over single-tier drones due to the flexibility of selecting different altitudes for drones.   Also, very few works investigate the performance of drones in the presence of conventional cellular networks. None of them critically identifies the scenarios when the drone network may be of potential significance to boost the performance of traditional cellular networks. 

In the following section, we investigate the feasibility of a multi-tier drone architecture over single-tier drones in a variety of urban environments and identify the scenarios in which the multi-tier architecture may be useful over single-tier drone networks and terrestrial cellular network.

\section*{Feasibility of Multi-tier Drone-Aided Cellular Architecture}

We consider a terrestrial cellular network composed of BSs  and users distributed as a homogeneous PPP  with intensities $\lambda_t$ and $\lambda_u$, respectively. Each BS transmits using a fixed power $P_t$. We consider a two-tier drone network where the big drones follow a 2D homogeneous PPP $\Phi_m$ with intensity $\lambda_m$  at a fixed altitude $h_m$. The small drones follow a 2D homogeneous PPP $\Phi_{s}$ at an altitude $h_{s}$ with density $\lambda_{s}$.  All big drones transmit with power $P_m$ whereas the small drones transmit at power $P_{s}$. Both tiers of drones as well as the terrestrial BSs use the same spectrum for transmission.
We consider the maximum received signal power-based association for the users. 
If multiple users are associated to the same BS, each user gets a channel with equal probability.
We consider turning off the BSs or drones with no users associated to them.

\subsection*{Path-Loss Model}
\subsubsection*{Terrestrial Network}
We consider a terrestrial cellular network with distance-dependent path-loss model. The channel between a user and a terrestrial BS is subject to path-loss and small-scale fading. The channel power gain is proportional
to $h d^{-\alpha}$, where $d$ is the distance between the user and the BS, and $\alpha$ denotes the path-loss exponent. The small-scale fading power gain $h$ is exponentially distributed with unit mean.

\subsubsection*{Multi-tier Drone Network}
The RF signals generated by a drone first travels through the free space until they reach the man-made urban environment, where some additional loss (referred to as excessive path-loss) is incurred  due to foliage and/or urban environment. 
The excessive path-loss is  random in nature and cannot be characterized by a well-known distribution. As such, the mean value of excessive path-loss $\eta_{\epsilon}$  has been considered in  most of the existing studies.
It is noteworthy that the RF transmissions from a given drone fall into  three propagation
groups, Line-of-Sight (LoS) propagation, non-LoS (NLoS) propagation via strong reflection and refraction,
and a very limited contribution (less than 3\% as reported in \cite{MAG}) by the deep fading resulting from consecutive reflections and diffractions. As such, the third group has been discarded in most of the relevant research studies.
Since the excessive path-loss depends largely on the first two propagation group, $\eta_{\epsilon}$   can be considered to be a constant that can be obtained by averaging all samples in a certain propagation group. The values of $\eta_{\epsilon}$ are listed for various frequencies and urban environments in \cite[Table II]{MAG}.

The AtG path-loss can thus be defined as follows~\cite{OLA}:
\begin{equation}
\mathrm{PL}_{\epsilon}=\mathrm{FSPL}+ \eta_{\epsilon}
\end{equation}
where $\epsilon \in \{\mathrm{LOS}, \mathrm{NLOS}\}$ and free space path-loss (FSPL) can be evaluated using the standard Friis equation, i.e.,
$
\mathrm{FSPL}= 20 \mathrm{log}_{10} \left(\frac{4 \pi f_c d}{c}\right),
$
where $f_c$ is the carrier frequency (Hz), $c$ is the speed of light (m/s), and $d$ is the distance between the drone and  the receiving user.

%The probability to have a LOS or NLOS connection depends on the propagation environment and can be approximated as follows~\cite{OLA}:
%\begin{equation}
%P_{\mathrm{LOS}}=\frac{1}{1+\mathrm{exp}[-b(\theta-a)]}
%\end{equation}
%where $\theta$ is the elevation angle between the drone and the served user (in degree).
%
%
%In drone networks, a ground user  experiences strong as well as weak reflected components 
%with different probability of occurrences based on environment, density and height of the buildings, and elevation angle of the drones. 

The probability of having LoS for user $i$ depends on the altitude of the drone-cell, $h_k$ where $k \in \{m,s\}$, and the horizontal distance between the drone-cell and $i^{th}$ user, which is $r_i=\sqrt{(x_D-x_i)^2 + (y_D-y_i)^2}$ for the $i^{th}$ user located at $(x_i,y_i)$ and the drone-cell located at $(x_D,y_D)$ with elevation $h_k$. The LoS probability is given by 
$
P_{\mathrm{LOS}}(h_k,r_i)=({1+a \mathrm{exp}(-b(\mathrm{arctan}(\frac{h_k}{r_i})-a))})^{-1}
$
where $\mathrm{arctan}(\frac{h_k}{r_i})$ is the elevation angle between the drone and the served user (in degree). Here $a$ and $b$ are constant values that depend on the choice of urban environment (high-rise urban, dense urban, sub-urban, urban). They are also known as S-curve parameters as they are obtained by approximating the LoS probability ({\em given by  International Telecommunication Union (ITU-R)} \cite{ITU,OLA}) with a simple modified
Sigmoid function (S-curve). Subsequently, the approximate LoS probability can be given for various urban environments while capturing the buildings' heights distribution, mean number of man made structures, and percentage of the built-up land area, of the considered urban environment.

The NLOS probability can then be defined as 
$
P_{\mathrm{NLOS}}(h_k,r_i)=1-P_{\mathrm{LOS}}(h_k,r_i).
$

%In this setting, the altitude of the user, and the antenna heights of both the users and the drone-cell are neglected. 
The path-loss expression can then be written as~\cite{E3D}
%\begin{equation}
$L(h_k,r_i)=20 \mathrm{log} \left(\sqrt{h_k^2+r_i^2}\right)+AP_{\mathrm{LOS}}(h_k,r_i)+B$
%\label{eq}
%\end{equation}
where $A = \eta_{\mathrm{LoS}}- \eta_{\mathrm{NLoS}}$, $B = 20 \mathrm{log}\left( \frac{4\pi f_c}{c} \right) + \eta_{\mathrm{NLOS}}$,
$\eta_{\mathrm{LoS}}$ and $\eta_{\mathrm{NLoS}}$ (in dB) are, respectively, the losses corresponding to the LoS and non-LoS reception depending on the environment.  
Table~II specifies the system parameters used for simulation results.

We would like to emphasize that the considered air-to-ground (AtG) propagation model is simple, general to capture various urban environments (such as high-rise urban, dense urban, sub-urban, urban), and has been applied in  various network settings to date. For instance, the considered AtG model has been used for the analysis of sophisticated altitude optimization and deployment of  static/mobile drones in several research studies  with the objective to either maximizing the coverage for users or to minimizing network delays or to minimize the power consumption of drones \cite{OLA, E3D, OTT, MIT, UAVU}.

\begin{table}
\caption{Simulation parameters}
\begin{tabular}{ |l|l|l| }
\hline
\multicolumn{2}{ |c| }{System parameter} & Value \\
\hline
\hline
\multirow{5}{*}{Environment}
&&$(a, b, \eta_{\mathrm{LOS}}, \eta_{\mathrm{NLOS}})$\\
\hline
 & Dense urban   & (12.08, 0.11, 1.6, 23) \\
 \hline
 & Sub urban & (4.88, 0.43, 0.1, 21) \\
 \hline
 & High rise urban & (27.23, 0.08, 2.3, 34) \\

\hline
    \multicolumn{2}{ |c| }{Carrier frequency ($f_c$)} & 2.5 GHz\\
    \hline
    \multicolumn{2}{ |c| } {Speed of light ($c$)} & $3\times 10^8$ m/sec\\
    \hline
     \multicolumn{2}{ |c| }{Path-loss exponent ($\alpha$)} & 4\\
      \hline
      \multicolumn{2}{ |c| }{Noise power ($N_0$)} & $1 \times 10^{-15}$ W/Hz\\
      \hline
      \multicolumn{2}{ |c| }{Intensity of terrestrial BSs ($\lambda_t$)} & 20\\
      \hline
      \multicolumn{2}{ |c| }{Intensity of drones ($\lambda$)} & 10\\
      \hline
      \multicolumn{2}{ |c| }{Altitude of big drones ($h_m$)} & 3000 m\\
      \hline
      \multicolumn{2}{ |c| }{Altitude of small drones ($h_s$)} & 150 m \\
      \hline
      \multicolumn{2}{ |c| }{Power of big drones ($P_m$)} & 40 W\\
      \hline
      \multicolumn{2}{ |c| }{Power of small drones ($P_s$)} & 5 W\\
      \hline
      \multicolumn{2}{ |c| }{Power of terrestrial BSs ($P_t$)} & 2 W\\
      \hline
    
\end{tabular}
\end{table}

\subsection*{Results and Discussions}
The spectral efficiency (SE) of transmission to a typical user can be defined as  $
R=\mathrm{log}_2(1+\mathrm{SINR})$, where
%\begin{equation}
SINR of a typical user $i$ associated to a drone BS $k \in \{m,s\}$ can be defined as :
$\mathrm{SINR}=P_k L(h_k,r_i)/[N_0+\sum_{j \in \Phi_t, j \neq k}P_t L(i,j) + \sum_{l \in \Phi_l, l \neq k}P_l L_k(h_l,r)]$, where $L(i,j)$ denotes the path-loss between terrestrial BS $j$ and user $i$, $l \in \{m,s\}$ and $N_0$ is the noise power.
%$\mathrm{SINR}=P_t L_{(\cdot)}/[N_0+\sum_{i \in \Phi_t, i \neq j}P_t L_i{(\cdot)} + \sum_{k \in \Phi_k, i \neq j}P_k L_k{(\cdot)}]$
%\end{equation}
%in which $N_0$ is the noise power and $L{(\cdot)}$ is the total path-loss due to distance and fading in terrestrial networks.
% and $L{(\cdot)}=L(h,r_i)$ for drone networks.  
%If $n$ users are associated to a BS, the spectral efficiency of transmission to each user becomes $\frac{1}{{n}}\mathrm{log}_2(1+\mathrm{SINR})$. 
\subsubsection*{Multi-tier vs Single-tier Drones}
\begin{figure}[t]
\begin{center}
\includegraphics[scale=.55]{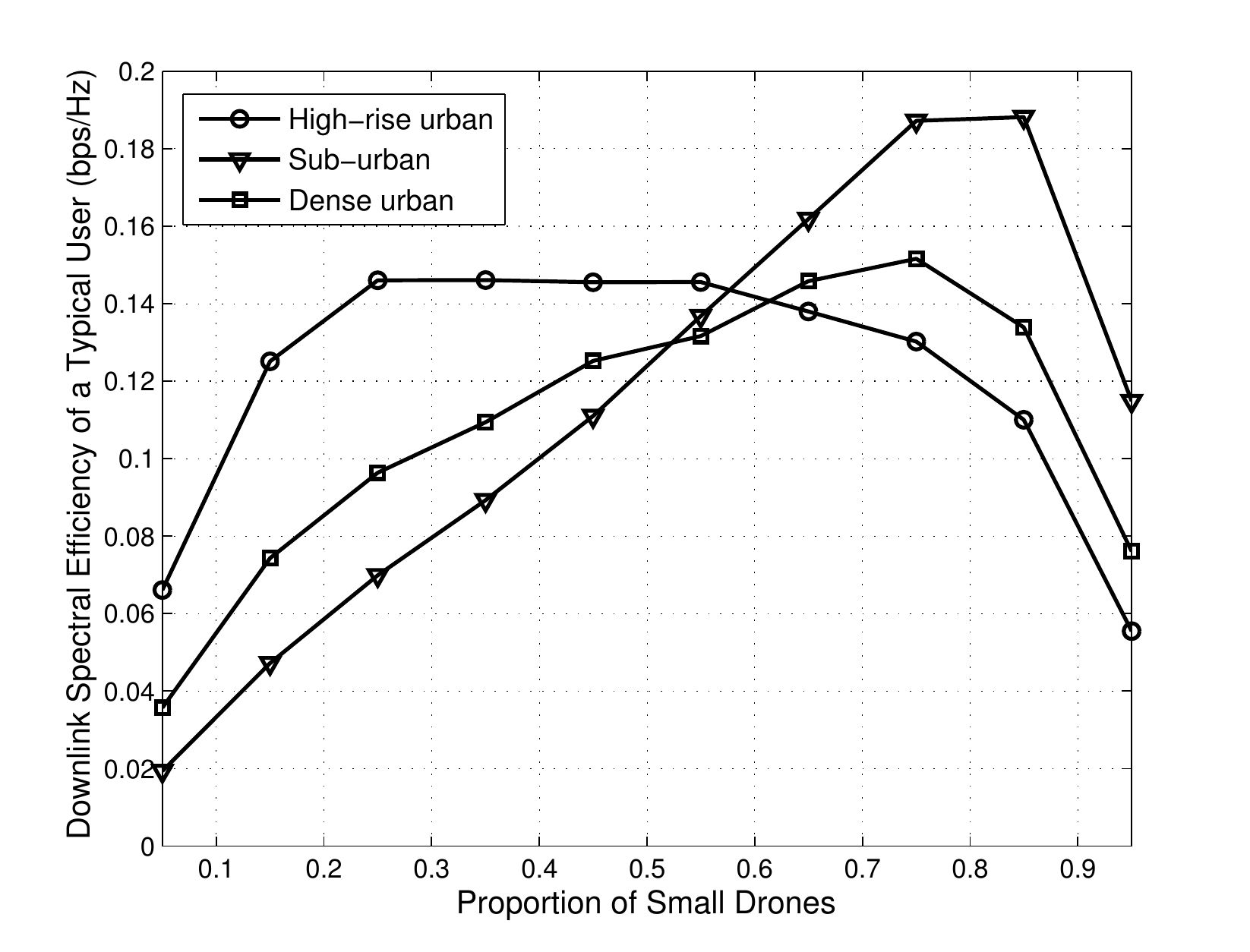}
\end{center}
\caption{ Downlink SE  of a typical user in multi-tier drone network with intensity of drones $\lambda$ = 10 as a function of the proportion of small drones (for $h_m = 3000$~m, $h_s = 150$~m).}
\label{env}
\end{figure}
From \figref{env}, we have the following observations:
\begin{itemize}
\item   The spectral efficiency  of transmission to a typical user with optimal proportion of big and small drones is higher than that with very large proportion of big  drones (i.e., similar to a single-tier of big drones) or small drones (i.e., similar to a single-tier of small drones). This is true for any of the chosen environments and demonstrates  the significance of multi-tier drones over single-tier drones. Big drones improve signal strength due to high power and  LoS connectivity; however, they are vulnerable to higher path-loss.  Small drones, on the other hand, increase signal strength by reducing path-loss. Nonetheless, their power and LoS connectivity is relatively less. It is thus beneficial to consider different types of drones to balance the trade-off between LoS connectivity and path-loss. 
\vspace{1mm}

{\em Remark:} In a high rise urban scenario, by choosing an optimal proportion, SE performance is observed to be improved approximately up to 133\%  compared to single-tier small/big drones. On the other hand, in dense urban scenario, multi-tier network improves the SE by 75\% and 250\% compared to small and big drone tiers, respectively.

\item Increasing the number of big  drones generally improves the signal quality due to high LoS probability and large transmission power. However, interference power is also high. The choice of optimal proportion of big drones in a multi-tier architecture is thus essential.

\item In most environments, a higher proportion of small drones is plausible. Nonetheless, in a high rise urban environment, a higher proportion of big drones needs to be deployed compared to other environments since they can provide higher LoS connectivity due to their high altitude. 

\item  If an optimal proportion of drones is selected, the performance of a typical drone user is better in a dense urban environment when compared to that in a high-rise environment. This is due to severe shadowing, reflections, and scattering in high-rise urban environments.

\end{itemize}

\begin{figure}[t]
\begin{center}
\includegraphics[scale=.6]{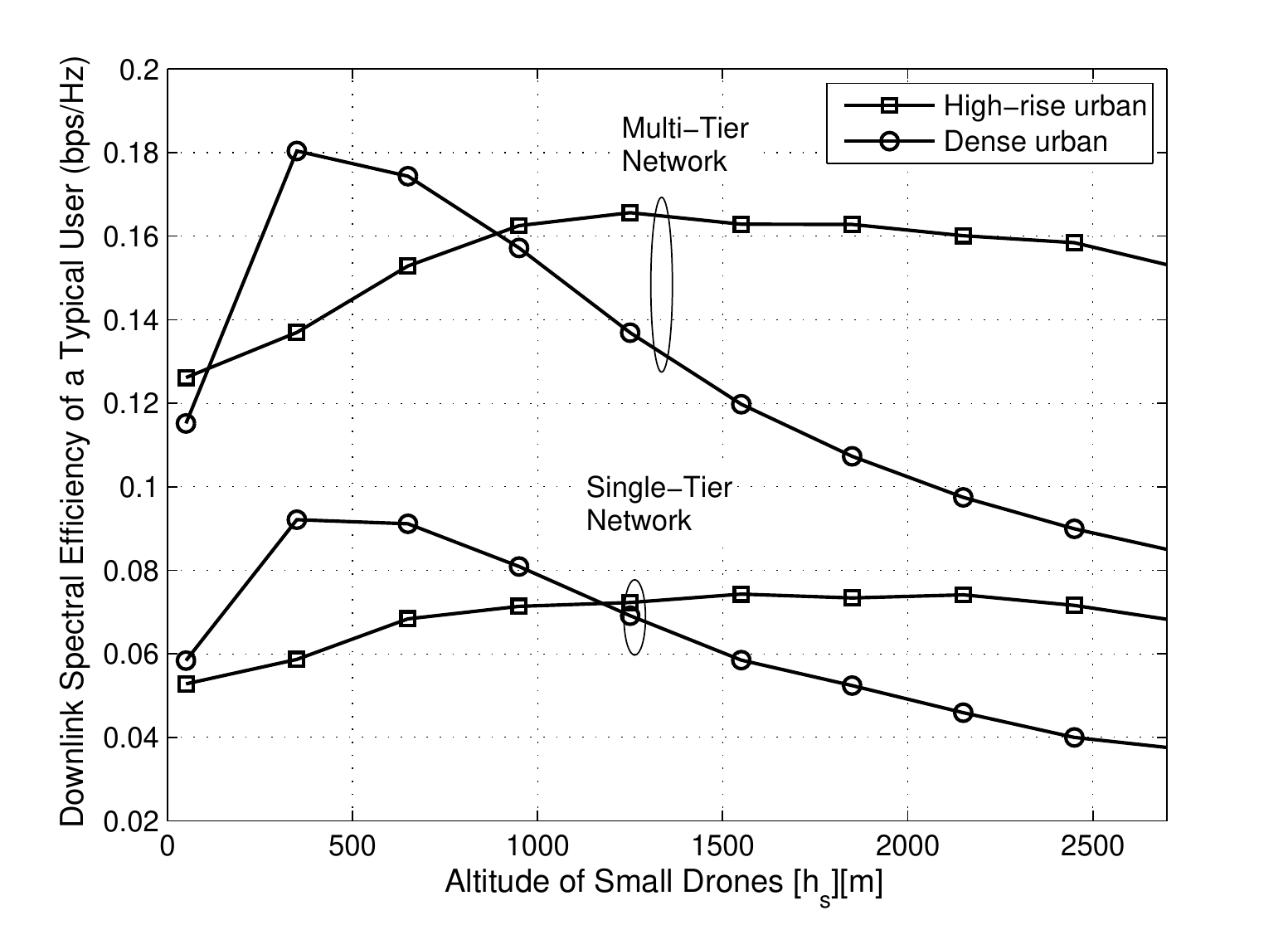}
\end{center}
\caption{Downlink SE  of a typical user in multi-tier drone network with intensity of drones $\lambda$ = 10 as a function of the altitude of small drones (for $h_m = 3000$~m).}
\label{alt}
\end{figure}

From \figref{alt}, we have  the following observations:
\begin{itemize}
\item With a proper choice of proportion of small drones (e.g., from \figref{env}), multi-tier drones improve the SE performance of a typical drone user  compared to single-tier drones for any chosen altitude of small drones. In a multi-tier setup, big drones improve the signal quality of a user due to high LoS probability and high power, but interference also increases in the network. A proper choice of the proportion of small drones  helps to balance the interference in the network and provide substantial gain over the single-tier network.
\item The optimal altitude to maximize the SE performance of a typical drone user  varies depending on the choice of urban environment. Due to poor LoS condition in a high-rise urban environment, small drones are better to be placed at higher altitudes than in a dense urban environment in order to improve the LoS connectivity and  therefore SE of transmission.
\end{itemize}

\subsubsection*{Multi-tier Drone-Assisted Terrestrial Cellular Networks}

\begin{figure}[t]
\begin{center}
\includegraphics[scale=.6]{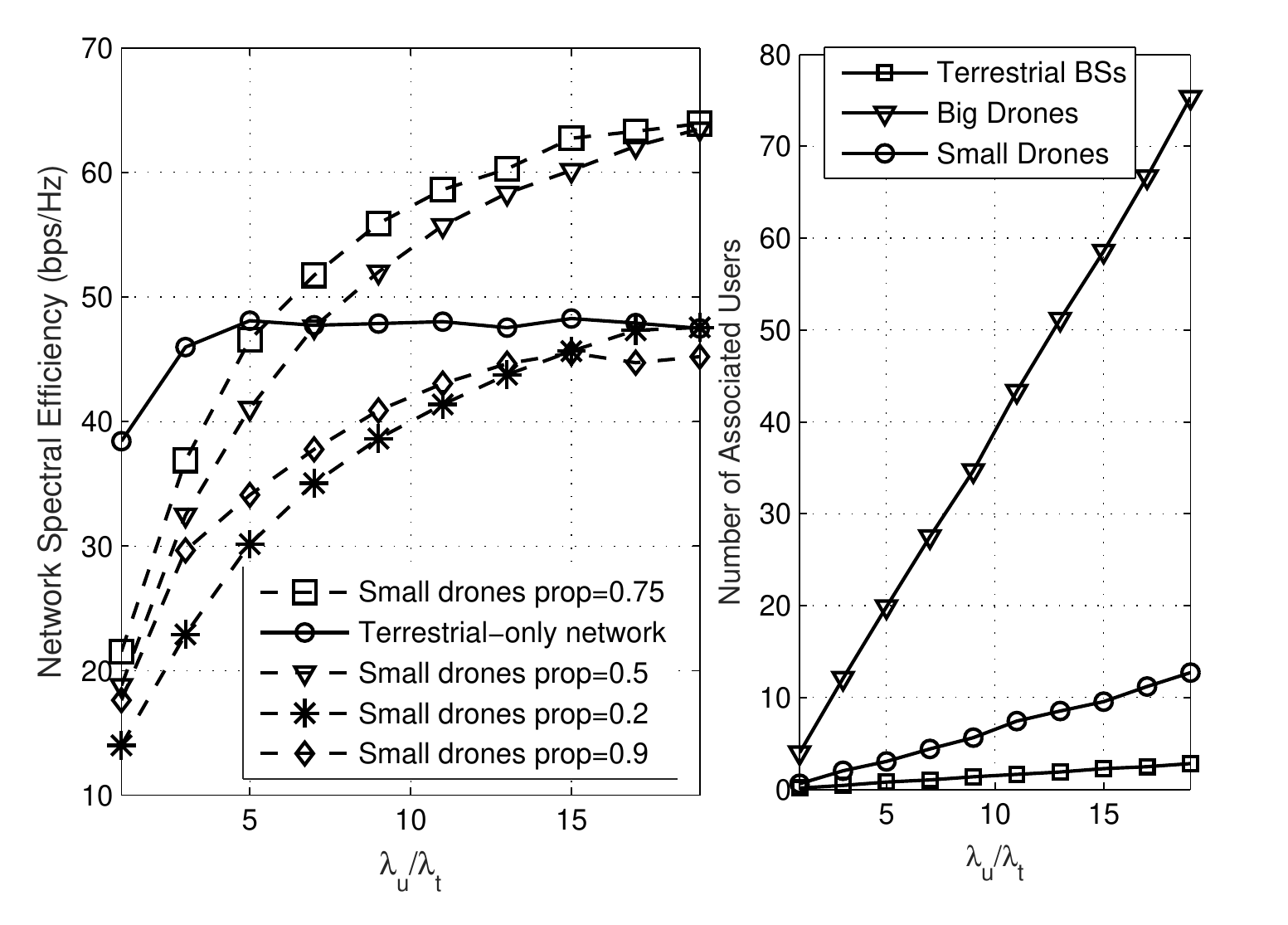}
\end{center}
\caption{(a) Network SE of a terrestrial network in the presence and absence of multi-tier drones,   as a function of  user to BS ratio $\lambda_{u}/\lambda_t$, $(\lambda_t=20)$, $P_t=2$~W, (b)  Number of users associated to terrestrial BSs and drones,  as a function of  user to BS ratio $\lambda_{u}/\lambda_t$ (for $\lambda_t=5$).}
\label{turnoff}
\end{figure}

In \figref{turnoff}, the network SE of terrestrial network with and without drones is compared where network SE denotes the sum of SE of all the users in the network. We plot network SE against $\lambda_{u}/\lambda_t$, which is defined as the proportion of the number of users in the network to the number of terrestrial BSs. 
We have the following observations:
\begin{itemize}
\item In high user to BS density scenarios (i.e., where the number of users is large compared to BSs), the multi-tier drone architecture improves the performance of a terrestrial network. The reason is that an increase in user intensity increases the traffic load per terrestrial BS and in turn reduces the share of transmission channel. With the incorporation of drones, the
probability of associating users to drones (as can also be observed from \figref{turnoff}(b)) increases since  users experience higher LoS probability.   Also, the load reduction in terrestrial BSs enhances the channel share of users that are associated to terrestrial BSs at the cost of increased interference from drones.  

\item  With the deployment of drones in correct proportion, the performance loss to the users (associated with terrestrial BSs) due to interference from drones is not significant.

\item  Multi-tier drone-aided cellular network  outperforms  terrestrial network if the proportion of small drones is chosen correctly. For example, if we choose the optimal point from \figref{env} which is for the drone-only network, significant performance gains can be observed. 
Also, under a co-channel deployment scenario, the performance of single-tier drones may not be beneficial over traditional terrestrial cellular networks as can be seen from the scenarios in which the proportion of small drone is 0.2 and 0.9 representing big drone tier and small-drone tier, respectively.
 
\end{itemize}

\subsection*{Extension to Multicast Systems}
The proposed architecture can be visualized
as a multicast system. Point-to-multipoint communication is typically achieved by providing
multiple connecting paths from a single location to multiple locations simultaneously. In
multicast systems, the system capacity increases linearly with the increase in the number of
receivers \cite{RA}. For example, the transmission rate of the multicast channel is $r$, then in the
presence of $K$ users, the achievable system capacity becomes $Kr$ \cite{RA}. Nevertheless, for wireless
multicast systems, the achievable system capacity becomes $Kr(K)$ where $r(K)$ is the multicast
transmission rate of the worst multicast user. As $K$ increases, $r(K)$ decreases because a typical
multicast transmission rate is adjusted to the worst-case user, i.e., the first term is increasing
while the second term is decreasing with $K$.
In the considered multi-tier drone architecture, each drone is connected to multiple users and
serves them. The users associated to the same drone BS can be referred to as multicast users.
The transmission rate of a multicast system is proportional to the worst performance user \cite{RA}.
However, in this article, we have considered the rate of a typical user (whose performance can be
referred as the average performance of users in a multicast group). Note that the altitude
of drone is much larger compared to the distance between the drone and worst multicast user.
As such, the average user rate does not vary much from the rate of the worst channel user. Thus,
by multiplying the spectral efficiency of a typical user with the number of multicast users $K$, we can obtain the throughput of the multicast system. For the considered PPP model, $K$ is a
random variable; therefore, the average value of $K$ can be used which is given in \cite{AM}.

%%%%%%%%%%%%%%%%%%%%%%%%%%%%%%%%%%%%%%%%%%%%%%%%%%%%%%%%

\section*{Conclusion and Future Directions}

We have studied the feasibility of multi-tier drones architecture over single-tier drones in terms of user spectral efficiency. We have investigated the impact of different urban environments on the optimal proportion of drones in a multi-tier drone network. To balance the impact of LoS probability and path-loss, incorporating drones in different tiers has been shown to improve the spectral efficiency of users. However, depending on the type of the urban environment, the optimal proportion of small and big drones may vary significantly. Not only the proportion but also the altitudes of drones in different tiers need to be optimized, e.g., small drones need to be placed at higher altitudes in high-rise environment compared to dense urban environment. We have also shown the performance gain of drone-assisted cellular network compared to the regular terrestrial network in terms of downlink spectral efficiency and have captured the network load conditions where the usage of drones can significantly outperform the traditional cellular network. 

Some possible future research directions are as follows:
\begin{itemize}
\item  Optimizing the  deployment of heterogeneous drones and their respective spectrum selection is crucial in order to minimize the deployment cost of drones along with their energy consumption.  

\item The use of unlicensed spectrum bands such as free space optics (FSO)  and mm-wave should be investigated under realistic propagation scenarios.  
Opportunistic spectrum selection (e.g., between mm-wave and RF) may be adopted to maximize the spectral efficiency performance.

\item Considering a more precise air-to-ground channel (AtG) model that incorporates temperature, wind, foliage,
near-sea environments, urban environments is crucial for a more precise performance analysis.
%Unfortunately, to the best of our knowledge, we are not aware of the availability of such a general
%AtG model that incorporates temperature, wind, foliage, near-sea environments. However, we
%believe that 
This is a potential future direction to work on.

\item Application specific optimization of deployment, mobility, and operation of drones (e.g., for online video streaming, multimedia broadcasting) in a converged broadband wireless networking scenario is an interesting research direction.

\item Instead of deploying drones by themselves, cellular network operators can share drones owned by some third parties (e.g.,  Amazon, Google). The economics of drone sharing will then need to be investigated  to optimize the network throughput considering the associated cost of  drone usage.

\item Integration of  UAVs to the Internet can potentially enable new IoT applications by leveraging cloud computing, web technologies, and service-oriented architectures. For such an  integrated environment,  UAV resources can be virtualized along with other network resources. Therefore, efficient methods will need to be developed for virtualization of UAV-enabled  5G networks.

\end{itemize}

\bibliography{IEEEfull,References}
\bibliographystyle{IEEEtran}

\begin{IEEEbiography}
%[{\includegraphics[width=1in,height=1.25in,clip,keepaspectratio]{S}}]
{Silvia Sekander}  received the BSc degree in
Computer Science and Engineering  from the
Bangladesh University of Engineering and Technology
(BUET), Bangladesh, in 2011, and the MSc
degree in Electrical and Computer Engineering
 from the University of Manitoba, Canada, in
2016. During her M.Sc. studies, she received the University of
Manitoba Graduate Fellowship (UMGF) and Manitoba Graduate Scholarship
(MGS). She is currently a Ph.D. student. Her
research interests include wireless communications
with a focus of resource management and
user association in heterogeneous networks including UAV-aided cellular networks. 

%Currently, she is working on
%Unmanned Aerial Vehicles (UAV)-aided wireless communications. During
%her undergraduate studies, she received the Dean?s List Award for her outstanding academic results at BUET. She also received the University of
%Manitoba Graduate Fellowship (UMGF) and Manitoba Graduate Scholarship
%(MGS) in her MSc program. She also worked as a software engineer
%for three years in Samsung Bangladesh R\&D Institute (SRBD), Dhaka.
\end{IEEEbiography}
\begin{IEEEbiography}
%[{\includegraphics[width=1in,height=1.25in,clip,keepaspectratio]{H}}]
{Hina Tabassum} received the B.E degree in electronic engineering from the N.E.D University of Engineering and Technology (NEDUET), Karachi, Pakistan, in 2004. She received during her under-
graduate studies the Gold medal from NEDUET and from SIEMENS for securing the first position among all engineering universities of Karachi. She then worked as lecturer in NEDUET for two years. In September 2005, she joined the Pakistan Space and Upper Atmosphere Research Commission (SUPARCO), Karachi, Pakistan and received there the
best performance award in 2009. She completed her Masters and Ph.D
degree in communications engineering from NEDUET in 2009 and King
Abdullah University of Science and Technology (KAUST), Makkah Province,Saudi Arabia, in May 2013, respectively. Currently, she is working as a post-doctoral fellow in the University of Manitoba, Canada. Her research interests include wireless communications with focus on interference modeling, spectrum allocation, and power control in heterogeneous networks.
\end{IEEEbiography}

\begin{IEEEbiography} 
%[{\includegraphics[width=1in,height=1.25in,clip,keepaspectratio]{ekram.jpg}}] 
{Ekram Hossain} (F'15)  is a Professor  in the Department of Electrical and Computer Engineering at University of Manitoba, Winnipeg, Canada. He is a Member (Class of 2016) of the College of the Royal Society of Canada. He received his Ph.D. in Electrical Engineering from University of Victoria, Canada, in 2001. Dr. Hossain's current research interests include design, analysis, and optimization of wireless/mobile communication networks, cognitive and green radio systems, and network economics. He has authored/edited several books in these areas (http://home.cc.umanitoba.ca/$\sim$hossaina). 
He was elevated to an IEEE Fellow ``for spectrum management and  resource allocation in  cognitive and cellular radio networks". He received the {\em 2017 IEEE ComSoc TCGCC (Technical Committee on Green Communications \& Computing) Distinguished Technical Achievement Recognition Award} ``for outstanding technical leadership and achievement in green wireless communications and networking". Dr. Hossain is an elected Member of the Board of Governors of the IEEE Communications Society for the term 2018-2020. 

\end{IEEEbiography}

%
%
%
%
%[{\includegraphics[width=1in,height=1.25in,clip,keepaspectratio]{H}}]{Hina Tabassum} received the B.E degree in electronic engineering from the N.E.D University of Engineering and Technology (NEDUET), Karachi, Pakistan, in 2004. She received during her under-
%graduate studies the Gold medal from NEDUET and from SIEMENS for securing the first position among all engineering universities of Karachi. She then worked as lecturer in NEDUET for two years. In September 2005, she joined the Pakistan Space and Upper Atmosphere Research Commission (SUPARCO), Karachi, Pakistan and received there the
%best performance award in 2009. She completed her Masters and Ph.D
%degree in communications engineering from NEDUET in 2009 and King
%Abdullah University of Science and Technology (KAUST), Makkah Province,Saudi Arabia, in May 2013, respectively. Currently, she is working as a post-doctoral fellow in the University of Manitoba (UoM), Canada. Her research interests include wireless communications with focus on interference modeling, spectrum allocation, and power control in heterogeneous networks.

\end{document}